\documentclass[twocolumn,aps,prb,amsmath]{revtex4}
\begin{document}
\title{On the ''effective exponent theory'' of the Coulomb Luttinger liquid}

\author{Yasha Gindikin and V. A. Sablikov}

\affiliation{Institute of Radio Engineering and Electronics,
Russian Academy of Sciences, Fryazino, Moscow District, 141190,
Russia}

\begin{abstract}
The ''effective exponent theory'', developed by Wang, Millis and Das
Sarma in [Phys. Rev. B \textbf{69}, 167101 (2004), Phys. Rev. B
\textbf{64}, 193307 (2001)], fails to calculate correctly the
dynamic correlators of Coulomb Luttinger liquid. Main drawbacks are
(i) cutting off the Coulomb potential by killing its long-range
component, (ii) the absence of non-contradictory procedure to
determine the 'scaling cutoff' and correlation functions for true
Coulomb interaction.
\end{abstract}

\maketitle

\section{Introduction}

Wang, Millis and Das Sarma (WMS) proposed~\cite{WMS} an 'effective
exponent' approach to calculate the dynamic correlation functions in
a Luttinger liquid with Coulomb interaction (CLL) and developed this
theory in recent publications~\cite{comment,wang02} opposing it to
our approach\cite{GS,plds,erratum} to the same problem. They claimed
that (i) their method is 'more elementary' than our one, but is
applicable in a wider energy range, (ii) both methods are equivalent
in describing the threshold behavior of charge-density wave (CDW)
structure factor and the dynamic spectral function (after a
technical error in deriving the spectral function in our
work\cite{GS} has been eliminated\cite{plds,erratum}).

In this Comment, we show that:

i) The 'effective exponent theory' fails to take adequately into
account the long-range nature of the Coulomb interaction and does
not correctly describe the dynamic correlation functions.

ii) WMS's method contains a concealed procedure of fitting to the
results of independent approaches and hence is not a self-sufficient
theory. Correspondingly, the range of applicability of WMS's method
is at least not wider than that of the reference approach.

iii) WMS's results\cite{comment} for the CDW structure factor and
spectral function disagree with our asymptotically exact
results~\cite{GS,plds,erratum}, and also contradict their own
earlier numerical calculations\cite{WMS}. In particular, the ratio
of two results for the spectral function $\rho(q,\omega)$ tends to
infinity in the threshold region of $\delta \equiv
\omega-\omega_{q-k_F}\to+0$ like $\exp[C|\ln\delta|^{3/2}]$, that is
faster than any power of $1/\delta$.

The theory of WMS would not probably have been deserving a special
discussion, if it had not involved an important physical problem of
Coulomb interaction effects in low-dimensional electron systems and
methods of its description.

Coulomb interaction effects are a rather complicated matter,
studying which requires a delicate approach. The brash
approximations can destroy the studied model so that it will be far
from the realistic physical system. The analysis of the WMS
theory\cite{WMS,wang02,comment} is quite instructive to illustrate
this. Their theory distorts strongly the Coulomb potential by
killing its long-range component that is of most interest.

\section{Coulomb interaction problem}
In principle, the Coulomb interaction problem is solved in the
spinless Luttinger liquid theory. The bosonization method allows one
to exactly calculate all the correlation functions for any type of
the electron-electron interaction potential, and, in particular,
does so for the long-ranged Coulomb interaction.\cite{Voit,Schulz}
As a result, correlators are expressed in terms of definite
integrals. Thus, the CDW structure factor $S_\mathrm{CDW}(q,\omega)$
equals\cite{LP,SG00}
\begin{equation}
S_\mathrm{CDW}(q,\omega)=\int_{-\infty}^{+\infty}dx\,e^{iqx}\int_{-\infty}^{+\infty}dt\,e^{i\omega
t}S(x,t)\;, \label{s1}
\end{equation}
with
\begin{eqnarray}
\nonumber
  S(x,t) &=& -\dfrac{1}{8\pi ^2}\partial_x^2\left(\exp[-v_F\int_{-
\infty}^{+\infty}\dfrac{dp}{\omega_p}\times \right.\\
   &&\left.(1-e^{-i\omega_pt-ipx})e^{-\alpha|p|}]\cos(2k_Fx)\right)\;.
\label{s2}
\end{eqnarray}
Here $v_F$ is the Fermi velocity, $\omega_q=|q|v_F/g(q)$ is bosonic
excitation spectrum, the interaction parameter $g(q)$ is related to
the Fourier-component of the interaction potential by
$g(q)=[1+V(q)/\hbar\pi v_F]^{-1/2}$.

Therefore, the task is reduced to the calculation of such integrals.
However, in the case of Coulomb interaction such calculation entails
serious mathematical difficulties that are connected inherently to
the slow decrease of the interaction potential at large distances.

True Coulomb interaction is well known to be a special one. The
distinctive feature of Coulomb interaction is that the integral of
Coulomb potential
$$V(x)\approx e^2/\sqrt{x^2+d^2}$$
diverges at large distance, $d$ being the quantum wire diameter.
Correspondingly, the Fourier-component of the potential goes to
infinity in the long-wave limit $q \to 0$ as
\begin{equation}
V(q)\sim 2e^2|\ln qd|\;. \label{pot}
\end{equation}
This manifests itself in the logarithmic divergency of the
collective excitation velocity $v(q)$ at $q \to 0$. Any
approximation of the Coulomb interaction by cutting off at large
distance loses this feature and hence makes the interaction a
non-Coulomb one.

The central question is whether long-ranged Coulomb interaction
leads to a qualitatively new behavior of 1D electrons as compared to
the short-ranged interaction case, which is easier to solve?

The classic paper of H.J. Schulz\cite{Schulz} gave the positive
answer to this question. Coulomb interaction was shown to have
dramatic consequences for the density-density correlations of a 1D
electron liquid. Specifically, the pair density correlator in the
Coulomb case decays with the distance in the extremely slow fashion
$\exp[-c(\ln x)^{1/2}]$, instead of a power law, typical of the
short-ranged interaction. In other words, in 1D systems Coulomb
interaction induces a quasi-long-range order, close to the Wigner
crystal.

Ref.~\onlinecite{Schulz} deals only with the static correlations.
Dynamic correlations are much more difficult matter because the
dependence on both $q$ and $\omega$ has to be addressed. Exact
results are apparently of key importance in the exploration of this
problem. They were not obtained until very recently.

The analytic approach to dynamic CLL correlators developed in our
paper\cite{GS} consists in the regular derivation of the integral
equation for $S_\mathrm{CDW}(q,\omega)$ and $\rho(q,\omega)$ and its
solution near the threshold. What is important is that no model
simplifications pertaining to the form of the Coulomb potential are
used in this approach.

In WMS's method\cite{comment,WMS,wang02}, no Coulomb integrals are
calculated analytically. WMS replaced the Coulomb interaction with
an 'effective' short-ranged interaction, i.e. cut off the
long-ranged Coulomb potential at large distance. Thus their approach
is a model one. But this is not its major shortcoming. The real
trouble lies in the arbitrariness of such replacement, as a
consequence of which the authors, turning to their method in
different publications\cite{comment,WMS,wang02}, each time obtain
different results. This becomes evident if one considers how Coulomb
interaction is transformed into short-ranged one. Since WMS in their
publications do not go into the details of such procedure, we
briefly outline it here for the case of the CDW structure factor
$S_\mathrm{CDW}(q,\omega)$.

\section{Effective exponent theory}

WMS start\cite{comment} with imposing the long-distance cutoff on
the Coulomb potential at some wave-vector $q_0$,
\begin{equation}
V(q)=2e^2|\ln((q_0+q)d)|\;. \label{screened}
\end{equation}
In contrast to Eq. (\ref{pot}), this potential does not diverge at
$q\to 0$, and hence is not the long-ranged Coulomb potential.

The structure factor $S_\mathrm{CDW}(q,\omega)$, having been
calculated with the screened potential of Eq. (\ref{screened}),
acquires the dependence on the cutoff parameter, i.e.
$S_\mathrm{CDW}(q,\omega)$ is transformed into
$S_\mathrm{CDW}(q_0;q,\omega)$. However, even for such cut off
interaction no formulas are derived. Instead, the authors refer to
common results for the short-ranged interaction, according to which
\begin{equation}
S_\mathrm{CDW}(q,\omega)\propto(\omega-\omega_{q-2k_F})^{\alpha_\mathrm{CDW
}}\;, \label{sshr}
\end{equation}
where the exponent equals $\alpha_\mathrm{CDW}=g(0)-1$.

WMS 'simply replace' the Coulomb interaction parameter
$g(q)\sim|\ln |qd||^{-1/2}$ by $g(q_0)$ and postulate that the
'effective' exponent equals
\begin{equation}
\alpha_\mathrm{CDW}(q_0)=g(q_0)-1\;. \label{alpha}
\end{equation}

Determining $q_0$ is the main trouble of WMS's approach. Indeed, no
consistent theory exists until the procedure to determine $q_0$ is
defined. Assuming $q_0$ constant (a fitting parameter) results in a
standard power-law dependence of the structure factor
$S_\mathrm{CDW}(q,\omega)$. For CLL, this corresponds neither to the
numerical calculations nor the expectations, based on exactly
solvable models. To obtain the formulas similar to those of
Schulz,\cite{Schulz} WMS subject $S_\mathrm{CDW}(q_0;q,\omega)$ to
further modification by introducing the explicit dependence of $q_0$
on $\omega$ and $q$. In other words, the cutoff parameter $q_0$ is
replaced by a function of two arguments $q_0(q,\omega)$. Notice that
introducing the cutoff $q_0$, which depends on $\omega$ and $q$, can
not be justified by any renormalization group arguments.

WMS propose no regular way to obtain the function $q_0(q,\omega)$.
Consequently, their method does not solve the Coulomb interaction
problem. In fact, $q_0$ is chosen from the intuitive conjectures,
differently for different correlators being calculated. In the
case of $S_\mathrm{CDW}(q,\omega)$, WMS postulate without
convincing justification that $q_0(q,\omega)$ is determined from
the following equation,
\begin{equation}
\omega_{q_0}=\omega-\omega_{q-2k_F}\;. \label{gcdw}
\end{equation}
In the case of $\rho(q,\omega)$, they state that $q_0(q,\omega)$
should be determined from other equation,
\begin{equation}
\omega_{q_0}=\omega-\omega_{q-k_F}\;.
\end{equation}
In doing so, they ignore the fact that in both cases
$q_0(q,\omega)$ enters into the same expression for the potential
in Eq. (\ref{screened}). As a result, the electron-electron
interaction potential turns out to be different for different
correlators. This 'scaling cut-off'
procedure\cite{WMS,wang02,comment} is seen to be physically
inappropriate.

The most important problem WMS confront further is how to obtain
the end result for correlation functions, even having the function
$q_0(q,\omega)$. In an earlier version of their
theory\cite{wang02}, $S_\mathrm{CDW}$ was determined according to
Eqs. (\ref{sshr}),(\ref{alpha}),(\ref{gcdw}), but this result did
not satisfy the authors, and in the final release\cite{comment}
they declared another procedure.

WMS postulated that the structure factor satisfies the following
differential equation (see Eq. (14) of Ref.~\onlinecite{comment})
\begin{equation}
\label{difur} \dfrac{\partial \ln
S_\mathrm{CDW}(\epsilon)}{\partial \ln
\epsilon}=\alpha_\mathrm{CDW}(q_0)\;,
\end{equation}
where $\epsilon$ denotes a normalized energy deviation from the
threshold, $\epsilon=(\omega-\omega_{q-2k_F})/\omega_0$, with
$\omega_0=v_F/\beta d$, $\beta=[\pi v_F/2e^2]^{1/2}$.

Eq. (\ref{difur}) is seen to be a fundamental principle of the
'effective exponent theory', which allows WMS to find the dynamic
structure factor analytically. However, no proof is provided that
the structure factor, given by Eqs. (\ref{s1}),(\ref{s2}), indeed
satisfies this equation, at least asymptotically in the threshold
region of $\epsilon \to 0$.

Solving Eq. (\ref{difur}), WMS obtain the following expression for
$S_\mathrm{CDW}$ (Eq. (15) of Ref. \onlinecite{comment}),
\begin{equation}
S_\mathrm{CDW}(q,\omega)\propto\exp(-2\beta|\ln\epsilon|^{1/2})/\epsilon\;.
\label{swms}
\end{equation}

The asymptotically exact expression for $S_\mathrm{CDW}(q,\omega)$
at $\epsilon \to0$ was found in our paper\cite{GS},
\begin{equation}
\label{ff3} S_\mathrm{CDW}(q, \omega)\sim
\dfrac{v_F}{\omega}\dfrac{e^{-4\beta|\ln
\epsilon|^{1/2}}}{\epsilon |\ln \epsilon|^{1/2}}\;.
\end{equation}

The comparison of the WMS result (Eq. (\ref{swms})) with our
result (Eq. (\ref{ff3})) via dividing one expression by another
\begin{equation}
\dfrac{S_\mathrm{WMS}}{S_\mathrm{GS}}\propto\exp(2\beta|\ln\epsilon|^{1/2})
\end{equation}
shows that in the most interesting threshold region $\epsilon\to+0$,
the ratio goes to infinity. This means that above results for the
CDW structure factor are not equivalent, in contrast to what WMS
state\cite{comment}. Possibly, one may invent some prescript to fit
the effective exponent expression to the correct result but this
hardly will be of interest.

Exactly the same steps are used in\cite{comment,wang02} to obtain
the dynamic spectral function. The latest presentation of the
effective exponent theory\cite{comment} does not contain the
explicit form of the spectral function. However, WMS's strategy, as
it is presented in Ref. \onlinecite{comment}, straightforwardly
leads to the following expression,
\begin{equation} \rho(q, \omega)\propto \exp[-\frac1{6\beta}|\ln
\delta|^{3/2}]\;, \label{wmsrho}
\end{equation} which gives the threshold behavior of the CLL
dynamic spectral function in the spinless case. Our result for the
spectral function\cite{plds,erratum} is derived below, see Eq.
(\ref{rho}). In the threshold region $\delta \to+0$, the ratio
\begin{equation}
\dfrac{\rho_\mathrm{WMS}}{\rho_\mathrm{GS}}\propto\exp[\frac1{6\beta}|\ln
\delta|^{3/2}]\;,
\end{equation}
goes to infinity faster than any power of $1/\delta$, which means
that the results are incompatible.

It is interesting to note that the results of analytic effective
exponent theory contradict WMS's earlier numerical calculations also
in the spinful case\cite{WMS}. Indeed, the analytic 'effective
exponent procedure' of Ref.~\onlinecite{comment} gives in this case
$\rho(q, \omega)\propto \exp[-\frac1{12\beta}|\ln \delta|^{3/2}]$,
whereas in Ref.~\onlinecite{WMS} it was found that $\rho(q,
\omega)\propto \exp[-\frac1{6\beta}|\ln \delta|^{3/2}]$. The ratio
of these expressions again goes to infinity in the threshold region
as $\delta\to+0$.

The above examples show convincingly that the 'effective exponent
theory' is a fitting to the independent (analytic or numerical)
results, rather than a justified self-sufficient procedure.
Consequently, the WMS claim\cite{comment} that their 'effective
exponent theory' applies over a much wider energy range than our
method is misleading. The effective exponent theory can not give
correct results out of the applicability range of independent
reference theory.

\section{On the fitting to numerical calculations}

As an independent reference theory in their earlier work\cite{WMS},
WMS used the numerical calculation in the frame of standard
Luttinger model. WMS chose the functional dependence of the cutoff
parameter $q_0$ and correlation functions by fitting to the
numerical results. For this purpose, the standard expressions for
the correlation functions of the Luttinger model with short-ranged
interaction were modified by introducing fitting parameters in the
exponents.

The shortcomings of such approach are evident. First, correlation
functions are of primary interest in the threshold regions, where
they either diverge or quickly go to zero; however, it is in these
regions that the accuracy of numerical calculations is least
trustworthy. Second, the fitting formulas do not follow from any
physical theory, but rather represent a formal mathematical
interpolation.

Moreover, one has serious grounds to doubt whether WMS's fitting to
numerical calculations is valid.

1) The formula for the electron Green function used in WMS's
calculations is incorrect. Specifically, Eqs. (4), (5) of Ref.
\onlinecite{WMS} contradict to the very definition of the Green
function. Green's function is defined in\cite{WMS} as
$G_r(x,t)\equiv\langle\psi_r(x,t)\psi^\dag_r(0,0)\rangle$, where $r$
refers to the right-moving ($r=+1$) or left-moving ($r=-1$)
particles. The Fourier-transform of the Green function
$G_r(q,\omega)$ possesses the fundamental property: at zero
temperature, $G_r(q,\omega)=0$ if $\omega<0$, the energy $\omega$
being measured from the chemical potential.\cite{AGD} WMS's results
(4),(5) show that the Green function of the left-moving particles
$G_\mathrm{r=-1}(x,t)$ is analytic in the \emph{upper} half-plane of
the complex time-variable. This immediately gives that Green's
function $G_\mathrm{r=-1}(q,\omega)\ne 0$ at $\omega<0$, but instead
$G_\mathrm{r=-1}(q,\omega)=0$ at $\omega>0$. Thus analytic and
spectral properties of the electron Green function are violated in
WMS's work\cite{WMS}.

2) Fitting procedure uses the behavior of the spectral function
$\rho(q,\omega)$ at frequencies much larger then the Fermi energy,
where the Luttinger model is not justified. Thus, they introduce
some 'previously overlooked energy scale' $\omega_s$ that is defined
as $\omega_s=50(v_F/d)\sqrt{V_0}\exp(1/V_0)$, with $V_0=4e^2/\pi
v_F$ being a dimensionless measure of the interaction strength. The
spectral function is considered at $\omega >\omega_s$. Note that
$\omega_s$ is of several orders of magnitude greater than the
electron Fermi energy, which is far beyond the scope of the
exploited approach!

3) The expression for the quasi-1D Coulomb interaction potential is
incorrect in the high-momentum region, where analytical results were
fitted to numerical ones. WMS took
\begin{equation}
V(q)=2e^2\ln\left[\dfrac{\Lambda+q}{q}\right]\;, \label{wmspot}
\end{equation}
with the short-range cutoff $\Lambda\approx2.5/d$, $d$ being the
diameter of the quantum wire, see Eq. (2) of Ref. \onlinecite{WMS}.
WMS denote the short-range cutoff also by $q_0$, but here we reserve
$q_0$ for the long-range cutoff.  WMS claim that the potential
should be '$1/q$ for $q$ larger than some scale $\Lambda$ set by the
geometry and the wave function size'.\cite{WMS} Eq. (\ref{wmspot})
indeed guarantees that.

However, such behavior is incorrect, since the potential should
behave as $1/q^2$ in the high-momentum region \cite{Kramer}. The
correct formula is derived in Appendix A,
\begin{equation}
V(q)=e^2\ln\left[\dfrac{\Lambda^2+q^2}{q^2}\right]\;. \label{corpot}
\end{equation}
It describes adequately the behavior of the quasi-1D Coulomb
potential both in the long-wave ($V(q)\approx 2e^2|\ln qd|$) and in
the short-range ($V(q)\approx e^2 \Lambda^2/q^2$) regions.

Generally, WMS uncritically overestimate their results pertaining to
the region of high $q$ and $\omega$.

Thus, using Eq. (\ref{wmspot}), WMS found the strange behavior of
the plasmon frequency $\omega_q$ at $q\gg d^{-1}$, namely
$\omega_q\sim qv_F+{\rm const}$. This is not a unique property of
CLL, as they assert, but a trivial mistake. Correct Eq.
(\ref{corpot}) gives $\omega_q\sim qv_F$, without any additional
constant.

They also discovered some dramatic modifications of $\rho(q,\omega)$
at $\omega\gg\omega_s$ as compared to the short-ranged interaction
case due to 'the slow ($1/q$) decay of the Coulomb interaction in
the large momentum region'. The passion fades if one remembers that
Eq. (\ref{wmspot}) for Coulomb potential is incorrect, and
bosonization does not work there.

\section{The analytic approach to CLL}

In our paper\cite{GS}, we have developed an analytic method to
determine the exact behavior of the dynamic correlation functions
of the Coulomb Luttinger liquid in the vicinity of the threshold.
We have taken into account that the velocity $v(q)$ of bosonic
excitations in a CLL diverges as $q\to0$ and effectively used this
fact. We will now briefly describe it to show how to obtain the
correct expression for the spectral function on a regular basis.

As it is shown in\cite{plds}, the $r$-fermion dynamic spectral
function satisfies the following integral equation
\begin{equation}
\rho(q,\omega)=\dfrac{v_F}{4\omega}\int_{-\infty}^{+\infty}dQ\,\mathcal{K}(q-Q)\,
\rho(Q,\omega-\omega_{q-Q})\;, \label{rho_eq}
\end{equation}
where the kernel is $\mathcal{K}(\xi)=\left[r-g^{-1}(\xi){\rm
sign}(\xi)\right]^2$, the wave vector $q$ being measured from the
Fermi wave vector. To be specific, consider right-moving
particles.

The conservation laws impose the following restriction on the
spectral function: $\rho(q,\omega)=0$ at $\omega<\omega_q$; thus
$\rho(q,\omega)$ has a threshold at
$\omega\to\omega_q+0$.\cite{GS} Applying this argument to the RHS
of Eq.~(\ref{rho_eq}), we find that the integrand is nonzero in
two narrow bands of the width $\delta/d\sqrt{|\ln\delta|}$, where
$\delta=(\omega-\omega_{q})/\omega_0$, situated near to $Q\approx
0$ and $Q\approx q$. The direct calculation shows that the
contribution of the band near to $Q\approx q$ to the integral is
smaller than the one from the band $Q\approx 0$ by the factor of
$\delta$. Therefore, considering the asymptotic behavior of
$\rho(q,\omega)$ at $\delta\to 0$, we can retain only the region
of $Q\approx0$. On these grounds, we expand the RHS of Eq.
(\ref{rho_eq}) in the powers of $Q$, appearing in the combination
$(q-Q)$,
\begin{equation}
\rho(q,\omega)=\dfrac{v_F}{4\omega}\int_{-\infty}^{+\infty}dQ\,
\left[1-\dfrac{{\rm sign}\,q}{g(q)}\right]^2
\rho(Q,\omega-\omega_q)+\dots \label{expansion}
\end{equation}
In the case of Coulomb interaction, it is sufficient to consider
only the first term of the expansion in the RHS of
Eq.~(\ref{expansion}) to obtain the correct asymptotic behavior of
$\rho(q,\omega)$ at $\delta \to +0$.\cite{GS,plds,repl}

On the RHS of Eq.(\ref{expansion}), the integral
$\int_{-\infty}^{+\infty}dQ\rho(Q,\omega-\omega_q)$ equals, up to
a factor, the tunnelling density of states $N(\omega-\omega_q)$,
which has a soft gap in the Coulomb interaction case.
Consequently, our method shows that the spectral function
$\rho(q,\omega)$ turns to zero as $\omega\to\omega_q$, i.e. it
contains a pseudogap. The functional dependence of the tunnelling
density of states can be obtain either via direct integration, or,
more elegantly, by solving the integral equation, similar to
(\ref{rho_eq}).

Finally, in the limit $\delta\to0$ we obtain \cite{erratum,plds}
\begin{equation}
\rho(q, \omega)\sim \frac{A}{\omega}\dfrac{|\ln
\delta|^{1/2}}{\delta} \exp[-\frac1{3\beta}|\ln \delta|^{3/2}]\;,
\label{rho}
\end{equation}
where $A(q)=\left[1-g^{-1}(q){\rm sign}(q)\right]^2$. The obtained
form of the spectral function is consistent with numerical and
experimental results\cite{Dardel,Zwick,Kurmaev,Kawakami} without
any mystique.

\section{Conclusion}

To summarize, our theory describes exactly the threshold behavior
of the CLL correlators. In contrast, the 'effective exponent
theory'\cite{comment} strongly distorts the interaction, making it
neither short-ranged nor Coulomb one. This is the main reason of
the discrepancy between our results.

\section*{Acknowledgements}

This work was supported by RFBR (02-02-16953, 01-02-97017), INTAS
(01-0014), Russian Ministry of Science, the programs of Russian
Academy of Sciences "Low-dimensional quantum structures" and
"Strongly correlated electrons in semiconductors, metals,
superconductors and magnetic materials", and Russian program
"Integration" (I 02908/1040).

\appendix
\section{Quasi-one-dimensional Coulomb interaction potential}
Here we derive the interaction potential of electrons, confined in
a 1D quantum wire and interacting via Coulomb long-ranged forces.
The pair interaction potential equals

\begin{equation}
U(x_1,x_2)=e^2\int
G(\textbf{r}_{1}-\textbf{r}_{2})\rho(\textbf{r}_{1\perp})\rho(\textbf{r}_{2\perp})
d\textbf{r}_{1\perp}d\textbf{r}_{2\perp}\;. \label{u}
\end{equation}
The $x$-axis is directed along the wire,
$\textbf{r}_{i}\equiv(x_i,y_i,z_i)$,
$\textbf{r}_{i\perp}\equiv(y_i,z_i)$. $\rho(\textbf{r}_{\perp})$
is the transverse density distribution in the wire, with only
first subband populated. The transverse distribution is supposed
to be localized on scale of $r_{\perp}\approx d$. The Green
function is
\begin{equation}
G(\textbf{r})=\frac1{|\textbf{r}|}\;,\;G(\textbf{k})=\frac{4\pi}{k^2}\;.
\end{equation}
Denote $x=x_1-x_2$. The long-distance ($x\to\infty$) behavior of the
potential can be well approximated by $U(x)\approx
e^2/\sqrt{x^2+d^2}$. Although this simple approximation leads to
correct logarithmic behavior of the Fourier-transformed potential in
the long-wave region of $q\to 0$, it is insufficient to obtain
correct expression in the short-range region. Below one finds a
general derivation.

Represent the Green function in Eq. (\ref{u}) via the Fourier
integral to get
\begin{equation}
U(q)=e^2\int\dfrac{dxd\textbf{k}}{2\pi^2k^2}e^{iqx}
e^{-i\textbf{k}(\textbf{r}_1-\textbf{r}_2)}\rho(\textbf{r}_{1\perp})\rho(\textbf{r}_{2\perp})
d\textbf{r}_{1\perp}d\textbf{r}_{2\perp}.
\end{equation}
Performing integration w.r.t. $x$ gives $\delta$-function, which
eliminates integration w.r.t. $k_x$. Integration w.r.t.
$\textbf{r}_{i\perp}$ yields
\begin{equation}
U(q)=e^2\int\dfrac{d\textbf{k}_{\perp}}{\pi}\dfrac{1}{q^2+\textbf{k}^2_{\perp}}
|\rho(\textbf{k}_{\perp})|^2\;.
\end{equation}
For cylindric wire this becomes
\begin{equation}
U(q)=e^2\int_0^{\infty}\dfrac{2kdk}{q^2+k^2}|\rho(k)|^2\;.
\end{equation}
Use the fact that $\rho(k)$ has a scale $\Lambda\approx1/d$, such
that $\rho(k)$ is small at $k>\Lambda$. Neglecting for simplicity
the $k$-dependence of $\rho(k)$ for $0<k<\Lambda$, one gets
\begin{equation}
U(q)=e^2\int_0^{\Lambda}\dfrac{2kdk}{q^2+k^2}=e^2\ln\left[\dfrac{\Lambda^2+q^2}{q^2}\right]\;.
\end{equation}
At $qd\ll1$, this gives the logarithmic divergency
\begin{equation}
U(q)\approx 2e^2\ln\left[\dfrac{1}{qd}\right]\;.
\end{equation}
When $qd\gg1$,
\begin{equation}
U(q)\approx e^2\dfrac{\Lambda^2}{q^2}\;.
\end{equation}

\end{document}